\documentclass[journal,onecolumn]{IEEEtran}%
\usepackage{eurosym}
\usepackage{amsmath}
\usepackage{amsthm}
\usepackage{amsfonts}
\usepackage{amssymb}
\usepackage{graphicx}
\usepackage[usenames,dvipsnames]{color}
\usepackage{booktabs}%
\providecommand{\U}[1]{\protect\rule{.1in}{.1in}}

\theoremstyle{plain}

\theoremstyle{definition}

\newlength{\drop}

\newcommand*{\titleUL}
{\begingroup
\drop=0.1\textheight
\vspace*{0.5\drop}
\begin{center}
{\LARGE\textsc{$\quad$}}\\[0.5\drop]
{{\LARGE IEEE Access} \\ \vspace*{0.4cm} -- accepted for publication --}\\[\drop]
\rule{\textwidth}{1pt}\par
\vspace{0.5\baselineskip}
{\huge\bfseries Sparse Signal Processing Concepts \\ \vspace*{0.4cm} for Efficient 5G System Design
}\\[0.5\baselineskip]
\rule{\textwidth}{1pt}\par
\vfill
{\Large\textsc{Gerhard Wunder$^1$, Holger Boche$^2$, Thomas Strohmer$^3$, Peter Jung$^4$}}
\vfill
$^1$Technische Unversit{\"a}t Berlin\footnote{Gerhard Wunder is also with the Fraunhofer Heinrich-Hertz-Institut, Berlin.} (\textit{gerhard.wunder@hhi.fraunhofer.de})\\
$^2$Technische Unversit{\"a}t M{\"u}nchen (\textit{boche@tum.de})\\
$^3$University of California, Davis (\textit{strohmer@math.ucdavis.edu})\\
$^4$Technische Unversit{\"a}t Berlin (\textit{peter.jung@mk.tu-berlin.de})
\vspace{1cm}
\begin{abstract}
As it becomes increasingly apparent that 4G will not be able to meet the emerging demands of future mobile communication systems, the question what could make up a 5G system, what are the crucial challenges and what are the key drivers is part of intensive, ongoing discussions. Partly due to the advent of compressive sensing, methods that can optimally exploit sparsity in signals have received tremendous attention in recent years. In this paper we will describe a variety of scenarios in which signal sparsity arises naturally in 5G wireless systems. Signal sparsity and the associated rich collection of tools and algorithms will thus be a viable source for innovation in 5G wireless system design. We will also describe applications of this sparse signal processing paradigm in MIMO random access, cloud radio access networks, compressive channel-source network coding, and embedded security. We will also emphasize important open problem that may arise in 5G system design, for which sparsity will potentially play a key role in their solution.
\end{abstract}
\vfill
\vfill
\textbf{Keywords}: Compressed Sensing, Cloud Radio Acess Networks, Massive Random Access, Embedded Security, Source Coding
\vfill
{\itshape \copyright 2012 IEEE. Personal use of this material is permitted. Permission from IEEE must be obtained for all other
uses, in any current or future media, including reprinting/republishing this material for advertising or promotional purposes, creating new collective works, for resale or redistribution to servers or lists, or reuse of any
copyrighted component of this work in other works.}
\end{center}
\endgroup}

\graphicspath{{./figures/}}
\begin{document}

\date{June 13, 2014}

\pagestyle{empty}\titleUL\thispagestyle{empty}

\addtolength{\baselineskip}{0.1cm}

\newpage

\section{What drives 5G?}

\label{s:intro}

The introduction of 4G clearly marked a first peak of the smartphones'
revolution, offering high bandwidth mobile radio access almost anywhere and
anytime. However, as it becomes increasingly apparent that 4G will not be able
to meet the emerging demands of future mobile communication systems, the
question what could make up a 5G system, what are the key drivers, is still
open and part of intensive discussions. Actually with 5G and related visions,
we believe, mobile communications research is on the brink towards a new
innovation cycle \cite{Wunder2014_COMMAG} (see also www.metis2020.com, www.5gnow.de):

\begin{itemize}
\item The \emph{Internet of Things} (IoT) will connect billions of devices,
i.e., the things of our everyday life, which is far more than 4G can
technically and economically accommodate. This will then open up new ways to
monitor, assist, secure, control e.g. in the tele-medicine area, smart homes,
smart factory etc. In fact, the IoT could change the way we see the Internet
as a human-to-human interface towards a more general machine-to-machine platform.

\item \emph{Security, privacy, and data integrity} will be a key issue in the
5G market. Current security solutions e.g. for the IoT fall short due to the
sheer number of nodes which must be flexibly managed and distributed in the network.

\item Moreover, the \emph{Tactile Internet} (TI) comprises a vast amount of
real-time applications with extremely low latency requirements including
\emph{industrial wireless applications} such as \emph{Smart Grids}. Motivated
by the human tactile sense, which requires round-trip times in the order of
1ms, 5G can then be applied for steering and control scenarios implying a
disruptive change from today's content driven communications. This is far
shorter than current 4G cellular systems allow for, missing the target by
nearly two orders of magnitude.

\item \emph{Gigabit wireless connectivity} is required in large crowd
gatherings with possibly interactively connected devices using
angle-controlled 3D video streaming, augmented reality, etc.
\end{itemize}

These examples make it very clear that 5G networks are not only just about
providing higher rates for the next smartphone generation (although certainly
important!), but more about \emph{enabling, integrating services} and
\emph{embedded security} which both implement very different (virtually
contradicting) application requirements. From a technical perspective it seems
to be utmost challenging to provide such uniform service experience to users
under the premises of future heterogeneous networking and small cell
scenarios. Consequently, the radio access has to be \emph{flexible, scalable,
content aware, robust, reliable and efficient in terms of energy and
spectrum}. Actually, with the limitations of current 4G system, this would put
further pressure on the common value chains on which the operators rely in
order to compensate for investment costs for future user services. Hence,
there is a clear motivation for an innovative and disruptive re-design of
current mobile communication networks from scratch.

Having the short-comings (i.e., high latency, no embedded security, very bulky
control signalling architecture) of 4G in mind, we develop the elements of a
5G research agenda based on \emph{sparse signal processing}. Here, sparsity
typically means that only a few samples of the signal are actually non-zero
but their locations may not be known. This new paradigm has been an intriguing
topic in mathematics and signal processing in recent years. Note that
sparsity-based concepts have been succesfully applied in specific
communication problems, e.g. the \emph{peak power control problem} (see
\cite{Wunder2013_SPM}). We will argue that sparsity in communication signals
is a viable source for innovation in 5G wireless system design and will,
hence, appear as our basic methodology.

\section{Enabling 5G technical concepts}

\label{s:enabler}

Let us first identify a series of enabling 5G concepts and corresponding
research challenges which shall be approached with the new paradigm:

\begin{itemize}
\item[i)] \emph{Fast and scalable random access }is one 5G key concept to
handle the massive number of \emph{sporadic traffic} generating devices (e.g.
IoT devices, but also Smartphones' Apps etc.) which are most of the time
inactive but regularly access the network for minor updates with no human
interaction. Sporadic traffic will dramatically increase in the 5G market and,
obviously, cannot be handled with the bulky 4G random access procedures. Two
major challenges must be addressed to leverage successful 5G business
models:\ i) unprecedented number of devices asynchronously access the network
over a limited resource, ii)\ the same resource carries \emph{control
signaling and users' payload}. Dimensioning the channel according to classical
theory results in a severe waste of resources which, even worse, does not
scale towards the requirements of the IoT (low-cost, deep indoor coverage,
long life time of devices). Yet, since typically user activity
\cite{Zhu:sparseMU2011}, channel profiles \cite{Bajwa2010:CSchest} and message
sizes are compressible within a very large receive space, sparse signal
processing methodology is a natural framework to support sporadic traffic,
cf.~Sec. \ref{s:randomaccess} where we discuss a suitable "one shot" approach.

In addition, the TI requires ultra-fast acquisition in the order of 100$\mu s$
on the physical layer to enable the 1ms round-trip time
\cite{Wunder2014_COMMAG}. Notably, this implies that even small, say 1kBit
data bursts, result in a huge bandwidth requirement. Again, we will argue that
classical theory requires that \emph{for each real-time connected device} a
significant control signaling overhead is necessary to allow for swift channel
estimation, equalization and demodulation. Since, in addition, this traffic
class must be also extremely reliable, control signaling must be separated
from the data which is very inefficient and can be much better handled by
sparse signal processing.

\item[ii)] \emph{Densification of cells} together with \emph{cloud-powered
baseband processing} and \emph{wireless network virtualization} (so-called
\emph{Cloud-RAN}) is another 5G key concept to increase spectrum and energy
efficiency and handle the projected traffic growth \cite{Wunder2014_COMMAG}.
It is based on the deployment of many light base stations with overlapping
coverage, performing only signal conversion to/from the digital domain,
connected through a high capacity link to a cloud of data centers. Coordinated
processing of signals by multiple network nodes is a key design element in
such a virtualized cellular network. Yet, it is still mostly overlooked in the
literature that existing cooperative designs do not scale in terms of run time
requirements and required control information. In Sec. \ref{s:cloud} we
analyze existing schemes and discuss a new control signaling architecture
thereby efficiently exploiting not only the compressible channels but also the
number of effectively coordinated nodes (out of the total number of nodes). We
will also discuss the beneficial interacting role of sparse prediction and
coordination in this scenario.

\item[iii)] \emph{5G source coding concepts} for the massive number of
distributed sensors and actuators will be very different as well. Shannon's
famous separation theorem states that under appropriate conditions data
compression (source coding) and error protection (channel coding) can be
performed separately and sequentially, without any performance loss. While the
separation theorem has has tremendous impact on the design of communication
systems, in several practical scenarios the conditions of the Shannon's
separation theorem neither hold nor can be used as a good approximation.
Naturally, this raises the question whether and in which form \emph{Shannon's
separation theorem} holds true and can serve as a guiding design principle for
the communication scenarios we expect to encounter in 5G. In
Sec.~\ref{s:coding} we discuss the related concepts and algorithms.

\item[iv)] \emph{Security} will play a central role in 5G networks. In today's
communication systems there is an architectural separation between data
encryption and error correction. The encryption module is based on
cryptographic principles and views the underlying communication channel as an
ideal bit pipe. The error correction module is typically implemented at the
physical layer. It adds redundancy into the source bits in order to combat
channel impairments or multiuser interference and transforms the noisy
communication channel into a reliable bit pipe. While such a separation based
architecture has long been an obvious solution in most systems, a number of
applications have emerged in recent years where encryption mechanisms must be
embedded in the physical layer (\emph{called physical layer security/ or
embedded security }\cite{SchaeferBoche13SPM}). Embedded security is a
relatively new research area exploiting the \emph{stochastic and physically
unclonable }nature of the wireless channel including noise \emph{and} the
hardware, e.g. for symmetric key and fingerprint generation. It is a further
5G vision to embed security into the concepts of Sec.~\ref{s:randomaccess} --
Sec.~\ref{s:coding} from scratch and to explore the benefits and tradeoffs in
such innovative designs to enable scalable, fast security mechanism
implemented without user interaction, please see Sec.~\ref{s:security} for a
discussion of new approaches in this regard.
\end{itemize}

\begin{figure}[ptb]
\centering
\includegraphics[width=0.85\linewidth]{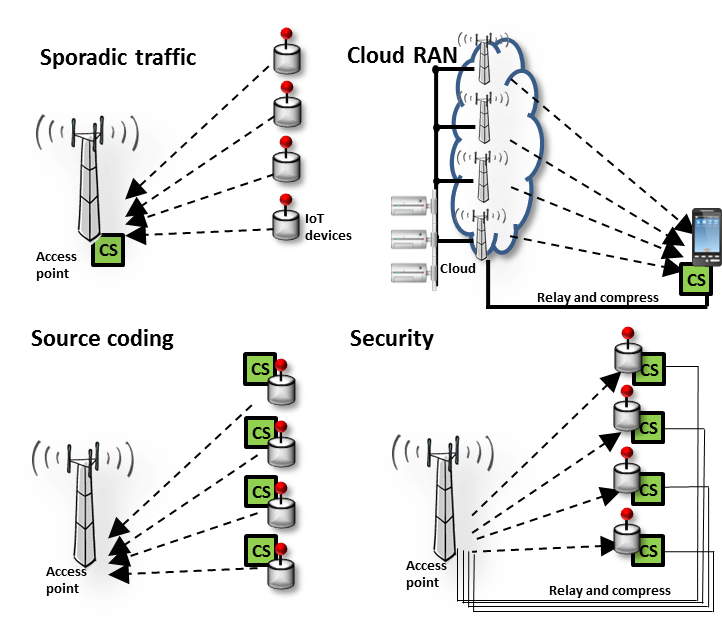} \vspace*{-1.7em}
\caption{Considered 5G deployment scenarios and respective "location" of
compressive sensing (CS) entities}%
\label{fig:scenarios}%
\end{figure}

The scenarios are depicted in Fig. \ref{fig:scenarios}.

\textbf{Notation}: For a vector $x\in\mathbb{C}^{n}$, $\lVert x\rVert_{p}$
denotes standard $\ell_{p}$--norm and $\lVert x\rVert_{0}$ counts the number
of non--zero elements in $x$. For sets/matrices we use capital/calligraphic
letters and $\text{vec}(X)$ denotes the vectorization of the matrix $X$. For a
given vector $x$, $\text{diag}(x)$ and $\text{circ}(x)$ refer to a diagonal
matrix with $x$ on its diagonal and to a circulant matrix (of appropriate
size) with $x$ as its first row, respectively. The nuclear norm $\lVert
X\rVert_{\ast}$ of a matrix $X$ is the $\ell_{1}$--norm of the vector of its
singular values.

\section{The Sparse Signal Processing Paradigm}

\label{s:sparsity}

At the core of compressive sensing (CS) lies the discovery, that it is
possible to reconstruct a sparse signal exactly from an underdetermined linear
system of equations \emph{and} that this can be done in a computationally
efficient manner via convex programming~\cite{FoucartRauhut2013}. Consider
$\Phi x=y$, where $\Phi$ is an $m\times n$ matrix of rank $m$ with $m<n$.
Here, $\Phi$ models the measurement (or sensing) process, $y\in{\mathbb{C}%
}^{m}$ is the vector of observations and $x\in{\mathbb{C}}^{n}$ is the signal
of interest. Conventional linear algebra wisdom tells us that in principle the
number of measurements $m$ has to be at least as large as the signal length
$n$, otherwise the system would be underdetermined and there would be
infinitely many solutions. Most data acquisition devices of current technology
obey this principle in one way or another.

Assume now that $x$ is $k$--sparse, i.e., $x$ satisfies $\Vert x\Vert_{0}\leq
k\ll n$ but we do \emph{not} know the locations of the non-zero entries of
$x$. Due to the sparsity of $x$ one could try to compute $x$ via an exhaustive
search, which however is NP-hard. Instead, the CS paradigm tells us that that
under certain conditions on the matrix $\Phi$ and the sparsity of $x$ we can
reconstruct $x$ from its measurements via linear or quadratic programming
techniques~\cite{FoucartRauhut2013}. While $x$ may not be sparse with respect
to the standard basis, in many cases we are dealing \emph{compressible}
signals which are well--approximated by signals which are sparse in some
specific domain (e.g. in the Fourier or wavelet domain). Thus, let $\Psi$ be a
$n\times n$ matrix which sparsifies $x$ and $y=\Phi\Psi x+z$, where $z$ is a
noise vector. To recover $x$ we consider the convex optimization problem:
\begin{equation}
\tilde{x}=\arg\min\lVert x\rVert_{1}\quad\text{s.t.}\;\lVert y-\Phi\Psi
x\rVert_{2}\leq\epsilon\label{eq:cs:l1min}%
\end{equation}
with $\epsilon$ depending on variance of $z$. Here, the $\ell_{1}$--norm
appears as the convex relaxation of $\lVert\cdot\rVert_{0}$.

A particularly important subject for the application in
multi-node/multi-terminal scenarios is the separation, respectively demixing
of multiple sparse signal contributions from the compressed superimposed
receive signal (\emph{compressive demixing}). Consider the model:
\begin{equation}
y=\Phi\sum_{p=1}^{P}\Psi_{p}x_{p}+e
\end{equation}
where each contribution $\Psi_{p}x_{p}$ is compressible in its own domain.
Here, the synthesis matrices $\{\Psi_{p}\}$ must essentially ensure that there
is no common intersecting subspace. Following \cite{McCoy:cdemix13} a
condition for exact (and/or stable) recovery is to avoid nontrivial
(well-separated) intersections between the per-user $\ell_{1}$-descent cones
at the true but unknown signals. See also \cite{Baraniuk2012_ARXIV} for a
quite general formulation of this problem.

In recent years, the idea of compressive sensing has been extended and
generalized in several ways. Adopting such a more general viewpoint (see
e.g.~\cite{Chandrasekaran2010}) we may assume that the objects of interest $x$
(for example $n$--dimensional vectors or $n\times n$--matrices) can be
well--approximated by superpositions $\sum_{j}c_{j}\psi_{j}$ of \emph{a few}
fixed atoms taken from an atomic set $\mathcal{A}\equiv\{\psi_{j}\}$. Compared
to the full linear span of \emph{all} atoms $\mathcal{A}$ the \emph{feasible
set of objects has then substantially reduced complexity}. In the standard CS
problem $\mathcal{A}$ corresponds to $\Psi$.

A more advanced set $\mathcal{A}$ is given by the superposition of a few
rank--one $n\times n$--matrices. The \emph{nuclear norm} $\lVert\cdot
\rVert_{\ast}$ will serve as convex relaxation of the rank--function, see
e.g.~\cite{CR08,Gross2011}. Assume that we observe the $n\times n$ matrix $x$
by taking $m$ Hilbert--Schmidt inner products of the form $(\Phi
X)_{l}:=\langle\Phi_{l},X\rangle$. In analogy to \eqref{eq:cs:l1min}, we can
attempt to recover $x$ via nuclear norm minimization, i.e., via:
\begin{equation}
\tilde{X}=\arg\min\lVert X\rVert_{\ast}\quad\text{s.t.}\;\lVert y-\Phi
X\rVert_{2}\leq\epsilon\label{eq:lmr:l1min}%
\end{equation}
which can be solved e.g.\ by semidefinite programming.

While by now we have a fairly good theoretical understanding of such low-rank
matrix recovery (LMR) problems as in~\eqref{eq:lmr:l1min}, only preliminary
results are known for the practical relevant combination of
\emph{simultaneously} low--rank and sparse structures, as present for example
in (sparse) \emph{compressive phase retrieval} or \emph{compressive blind
deconvolution} discussed in Sec. \ref{s:cloud} and Sec. \ref{s:randomaccess}.
Multi-objective ($\ell_{1}$ and nuclear norm) convex programs for
simultaneously sparse and low-rank matrices are limited by the so called
rank-sparsity incoherence and stable recovery cannot be achieved at the
theoretically optimal (for non--convex recovery) number of measurements
\cite{Oymak2012}.

In several applications we are confronted with nonlinear measurements, such as
intensity or quantized measurements. Thus, we may consider the more general
case where we get information about the object $x$ by taking $m$
non--adaptive, linear or nonlinear noisy observations of the form
$(y)_{l}=f(\langle\phi_{l},x\rangle)+(z)_{l}$. The famous \emph{phase
retrieval problem} falls in this setting, here $f(\langle\phi_{l},x\rangle) =
|\langle\phi_{l},x\rangle|^{2}$. Recently it has been
shown~\cite{candes:phaselift} that the phase retrieval problem can be casted
as a linear matrix--estimation problem under rank--constraints and exploiting
the LMR framework with further cone constraints, i.e., $\mathcal{A}$ is the
set of rank--one positive--semidefinite matrices. The approach in
\cite{candes:phaselift}, called PhaseLift, proceeds by \emph{lifting} the
absolute--square map on vectors to a linear map on rank-one matrices.
To be explicit, with $\Phi_{l}=\phi_{l}\phi_{l}^{\ast}$ and $X=xx^{\ast}$ one
has $(y)_{l}=\langle\Phi_{l},X\rangle+(z)_{l}=\left(  \Phi X\right)
_{l}+(z)_{l}$ which matches the LMR model, i.e., algorithms related to
\eqref{eq:lmr:l1min} can be used for recovery.
An even more constrained form of (noisy) phase retrieval problem is to recover
a \emph{sparse} complex vector from noisy intensity measurements. See
\cite{Ehler2013} for some recent theoretical results and further references.

\section{Compressive Multi-antenna Random Access}

\label{s:randomaccess}

In this section we will introduce a general \emph{compressive multiple antenna
random access}. In this model each device asynchronously accesses the network
thereby carrying overlapping data and control signals. In such a system, data
detectibility becomes increasingly erroneous the more control is interfering
with the data. Yet, the control must somehow interfere with data in order to
allow for (swift) estimation of the channel. This seems to be a contradicting,
irresolvable task at first sight. However, we show how we can cope with this
task by exploiting sparse signaling principles.

\subsection{Random access: A key application for sparse signal processing}

A key 5G application of the sparse signal processing paradigm is the evolution
of the random access channel (RACH) with asynchronous short--message support
which explicitly exploits compressibility. In RACH, sparse structures are
present in many directions: (i) due to user activity and the
near/far--behavior only a small but unknown subset of users participate in the
random access at a particular base station (ii) the mobile channels (its
spreading function) from the terminals to base stations are sparse in delay
and Doppler and (iii) MIMO channel matrices are often low--rank due to
collocation of antennas and (iv) sporadic traffic with short-message type
payload is intrinsically sparse.

On the other hand, multiple challenging tasks have to performed simultaneously
in such an improved compressive RACH architecture. Firstly, (i) the active
user set has to be identified. Then, (ii) the associated channel coefficients
for these users have to be estimated. The separate determination of the
channel characteristics is important for possible resource assignment for
successive high data rate uplink. Finally, (iii) the RACH data payload for
each active user has to be reconstructed. Without explicitly exploiting the
sparse structures discussed above, it seems to be practically impossible to
achieve steps (i),(ii), and (iii) in a single or few transmission steps.
Therefore, traditionally, only step (i) is accomplished in the RACH and steps
(ii) and (iii) are postponed to a synchronized uplink channel which comes then
with the already mentioned control overhead. Thus, \emph{exploiting the system
sparsity} can be a key enabler for IoT and, by similar reasons as we pointed
out in Sec. \ref{s:enabler}, for the TI on the physical layer.

Sparse reconstruction methods have been used here already in all these steps
separately. For example, step (i) user activation---also known as on-off
RACH---can be cast as a CS problem \cite{Fletcher09}, see also
\cite{Bockelmann:ETT2013}.
Multipath channel estimation is meanwhile a classical field for CS methods
\cite{Bajwa2010:CSchest}. Compressive demodulation or demixing of superimposed
signals, step (iii), is a further field of research
\cite{McCoy:cdemix13,Baraniuk2012_ARXIV}.

Let us discuss suitable detection strategies in the following. We will
distinguish between \emph{coherent concepts} and \emph{incoherent concepts}
where the receiver estimates the multi-antenna channel separately prior to
data demodulation or not, respectively.

\subsection{Coherent multiantenna receiver concepts}

Our target system to be investigated is described as follows: We adopt a
general model where multiple receive antennas $n_{r}$ are incorporated from
scratch. Transmission is on a frame by frame basis where the signal space
dimension $n$ within the frame can be very large, say several thousands
samples due to large bandwidth or large observation times. The time-space
(rows-columns) signal to be compressively sensed in some slot is given by the
matrix $Y\in\mathbb{C}^{n\times n_{r}}$. Let us assume that the channel
coherence time is essentially larger than the slot time. There are three
sources for sparsity or compressibility that should be exploited:

\begin{itemize}
\item \emph{Sporadic (Sparse) Traffic:} The communication in random access is
sporadic so that out of $n_{t}$ nodes only an unknown small subset of size
$k_{0}$ are actually active. Alternatively, we can assert certain
probabilities to each node. This is our \emph{primary source for sparsity}
within the receive space $\mathbb{C}^{n\times n_{r}}$.

\item \emph{Multipath Channels:} Communication is over a multipath channel
with delay spread $n_{d}$. For each $p$-th/$q$-th transmit/receive pair, the
channel vector $h_{q}^{p}$ contains the $n_{d}$ coefficients of the
\emph{channel impulse response} (CIR) which form the matrix:
\begin{equation}
H_{p}=\left[
\begin{array}
[c]{lllll}%
\vdots &  & \vdots &  & \vdots\\
(h_{1}^{p})_{i} & ... & (h_{q}^{p})_{i} & ... & (h_{n_{r}}^{p})_{i}\\
\vdots &  & \vdots &  & \vdots
\end{array}
\right]  \in\mathbb{C}^{n_{d}\times n_{r}} \label{eq:crach:pqchannel}%
\end{equation}
We assume that out of the $n_{d}$ channel coefficients in each column, only
$k_{1}$ in the CIRs are non-zero and the exact positions of the coefficients
within $H_{p}$ are unknown. This sparsity assumption is fulfilled in most
wireless communication scenarios and, indeed, channel estimation was one of
the first CS applications here \cite{Bajwa2010:CSchest}. Meanwhile, CS
estimation methods have been extended towards sparsity in the delay-Doppler
domain (see e.g. \cite{Tauboeck2010}). Large bandwidth channels tend to
exhibit sparsity only in the delay domain where the support of the pathes is
invariant though \cite{Cheng2013}. This is our \emph{second source for
sparsity}.

\item \emph{Compressible Short Messages:} Each user transmits a sequence
$x_{p}=C_{p}d_{p}$ where $d_{p}\in\mathcal{M}\cup\{0\}\equiv\mathcal{M}_{0}$
is from some modulation alphabet which, by the activity model includes a zero
energy symbol as well. The matrix $C_{p}\in\mathbb{C}^{n\times n}$ is some
designed or random code matrix. We assume that $k_{2}$ data symbols out of the
$n$ are actually non-zero. This is our \emph{third source of sparsity}. Note
that in the random case the transmitter has to be informed about the code
matrices Further reasoning for this assumption follows when taking intrinsic
compressibility of the data payload into account as it will be explained, for
example, in Sec. \ref{s:coding}.

\item \emph{Spatial corelations:} Another source of sparsity is the spatial
domain particularly when the number of antennas is large (the so-called
\emph{massive antenna regime}). In this situation the covariance matrix is
sparse but in some domain which is typically unknown; only in some special
cases this matric can be decomposed in the Fourier domain, e.g. for the linear
array \cite{Adhikar2013}. A natural question is then how to measure this such
(huge) covariance matrix or alternatively how to select a proper basis to
exploit the sparsity in the process. Additionally, this cannot be done
independent of the data detection process.

\item \emph{Topology}: Due to sparse connectivity, users will be separated by
their received transmission powers. This is e.g. a possible way to inherently
distinguish relevance of certain nodes in a cooperative design (see also the
Sec. \ref{s:cloud}).
\end{itemize}

As we outlined before, our goal is to achieve "one shot" transmission, i.e.,
user detection, channel estimation and data detection in one time slot.
Therefore, for each node $p$, a control sequence $s_{p}\in\mathbb{C}^{n}$
drawn from some pool of sequences known to the access, and an unknown data
sequence $x_{p}\in\mathcal{M}_{0}^{n}$ is transmitted. Each transmit sequence
has some individual power and the power is split between the control and the
data. The sampling equation for the receiver $q$ can be mathematically
expressed with \eqref{eq:crach:pqchannel} as\footnote{It is important to note
that this "underlay" signal model does not exclude the case where control and
data signalling are completely separated, e.g. by FDMA in the frequency
domain.}:%
\begin{equation}
y=\Phi\left(  \sum_{p=1}^{n_{t}}\left(  S_{p}+X_{p}\right)  H_{p}+Z\right)
=:\Phi(Y) \label{eq:crach:coherent}%
\end{equation}
whereby, for the sake of exposition, we use a circular matrix model
$S_{p}=\text{circ}(s_{p})$ and $X_{p}=\text{circ}(x_{p})$ to represent all
circular convolutions between transmit signals $s_{p}+s_{q}$ and channel
impulse responses $h_{q}^{p}$. The matrix $Z$ denotes additive white Gaussian
noise with variance $\sigma^{2}$ per component. Measurement on this matrix
signal $Y$ is performed with a linear mapping $\Phi$ giving the sample values
$y=\Phi(Y)$. If the same sampling is used independently for all receive
antennas, one could take also $\Phi\in\mathbb{C}^{m\times n}$ as the
measurement (compression) matrix. Compared with compressive demixing in
Sec.\ref{s:sparsity}, the dictionary matrices $\Psi_{p}$ are now related to a
circulant model. In CS for a single signal with a random measurement matrix
model $\Phi\Psi_{p}$ where $\Phi$ is isotropically--distributed this is
sometimes called the anisotropic case. Noteworthy, in the classical setting
this model resembles closely the overloaded multiple access channel which is
well understood. In the overloaded case, optimal \emph{mean squared error}
designs can be achieved \cite{Wunder2006_TSP} which is very different from the
compressive case considered here where such problems have not yet been touched.

The model in \eqref{eq:crach:coherent} is actually general enough to include
several recent models and once $H_{p}$ is assumed to be known or $X_{p}$ is
zero the determination of the remaining unknowns under sparsity assumptions
reduces (up to the anisotropy) to a standard CS problem. For example, if
$H_{p}$ is already known at the receiver, user activity and data demodulation
can be performed with CS--based multiuser detection methods,
see~\cite{Zhu:sparseMU2011} and further references in
\cite{Bockelmann:ETT2013}. However, in general $H_{p}$ itself has to be
estimated within the same transmission frame which is a challenging task both
from the algorithmic side as well es from the design point of view. Random
codebook and/or pilot assignments and spreading, respectively, could be a
possible strategy for this problem \cite{Bockelmann:ETT2013}. The limits of
such \emph{compressive demixing} strategies by convex methods has been
characterized for the random orientation model in \cite{McCoy:cdemix13} in a
quite general setting. According to \cite{McCoy:cdemix13} reliable demixing is
possible whenever the number measurements $m$ is sufficiently above the total
sum of all contributing statistical dimensions. In the sparse case the
statistical dimension of a single signal contribution amounts for the
effective dimension of $\ell_{1}$-descent cones at the unknown signal and
scales linearly with signal sparsity. A greedy approach to demixing for more
general signal manifolds is contained in \cite{Baraniuk2012_ARXIV}.

A different clever strategy could be the combination of \emph{illumination and
subsampling} as proposed for the random demodulator in \cite{Tropp:RandDemod}
and used for random access in \cite{WunderICC_2014}. Using the different
cancellation properties of stochastic i.i.d. data with deterministic pilots,
it is possible (i) to cumulate sufficient pilot power for sparsity--aware
channel estimation in a fixed window at the output of the random demodulator
and then (ii) to demodulate the whole data payload. To show the potential of
such approach let us consider the following 5G--like situation using LTE-A 4G parameters:

An LTE-A 4G frame consists of a number of subframes with 20MHz bandwidth; the
first subframe contains the RACH with one "big" OFDM symbol of $m=839$
dimensions located around the frequency center of the subframe. The FFT size
is $n=24578=24$k corresponding to the 20MHz bandwidth whereby the remainder
bandwidth outside PRACH is used for scheduled transmission in LTE-A, so-called
PUSCH. The prefix of the OFDM symbol accommodates delays up to 100$\mu s$ (or
30km cell radius) which equals 3000 dimensions. In the standard the RACH is
responsible for user aquisition by correlating the received signal with
preambles from a given set. Here, to mimic a 5G situation, we equip the
transmitter with the capability of sending information in "one shot", i.e., in
addition to user aquisition, channel estimation is performed and the data is
detected. For this a fraction of the PUSCH is reserved for data packets of
users which are detected in the PRACH. Please note the rather challenging
scanrio of only 839 subcarrier in the measurement window versus almost 24k
data payload subcarriers.

\begin{figure}[t]
\centering\includegraphics[width=\linewidth]{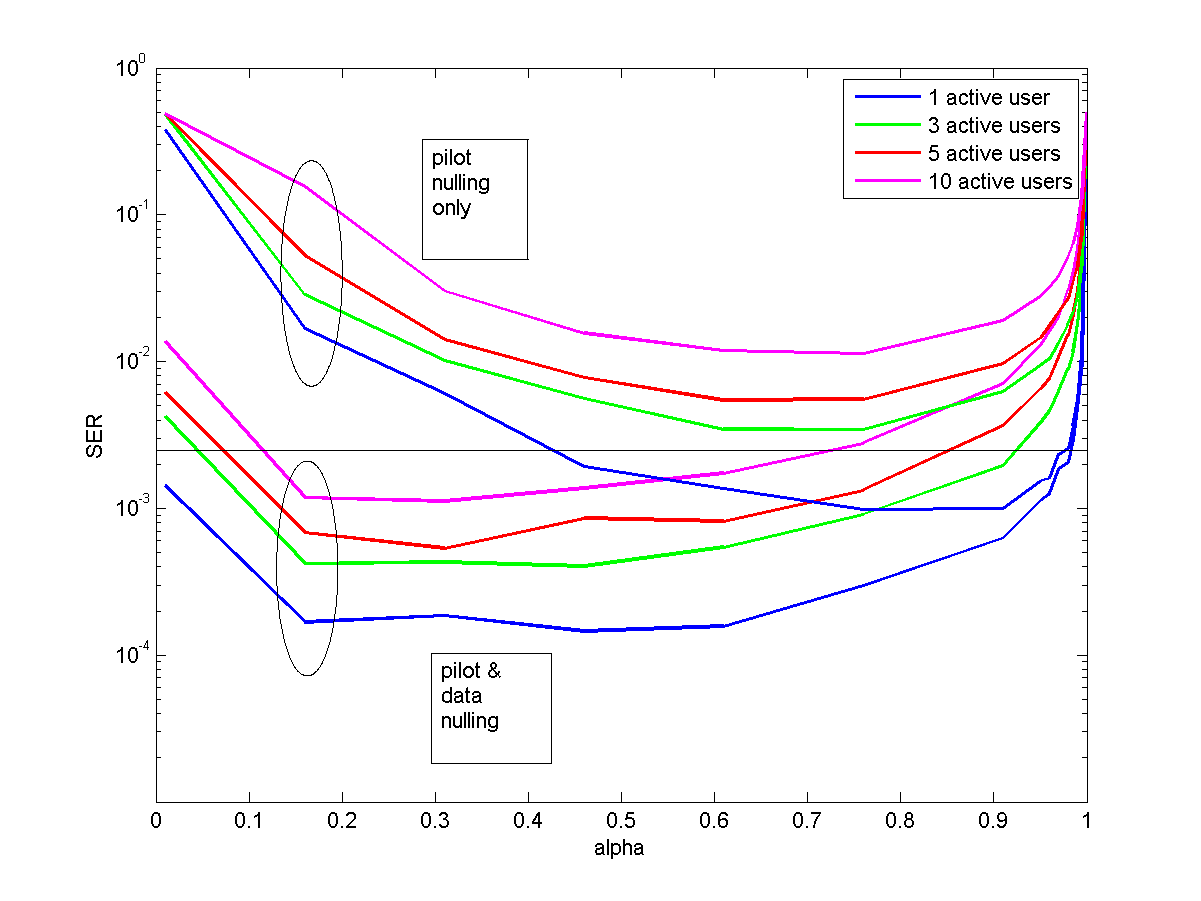}\caption{Averaged
BPSK SER in 5G \textquotedblright one-shot\textquotedblright\ random access in
a $20$MHz LTE-A standard setting at (overall) SNR=$20$dB. In the first setting
(upper curves), $m=839$ dimension out of $n=24576$ dimensions are used for CS.
This limits the control overhead to below 5\%. In the second case, pilots and
data are fully separated so that the performance is greatly improved at the
expense of a slightly increased control overhead ($<$14\%)}%
\label{fig:crach:ser}%
\end{figure}

\begin{figure}[t]
\centering\includegraphics[width=\linewidth]{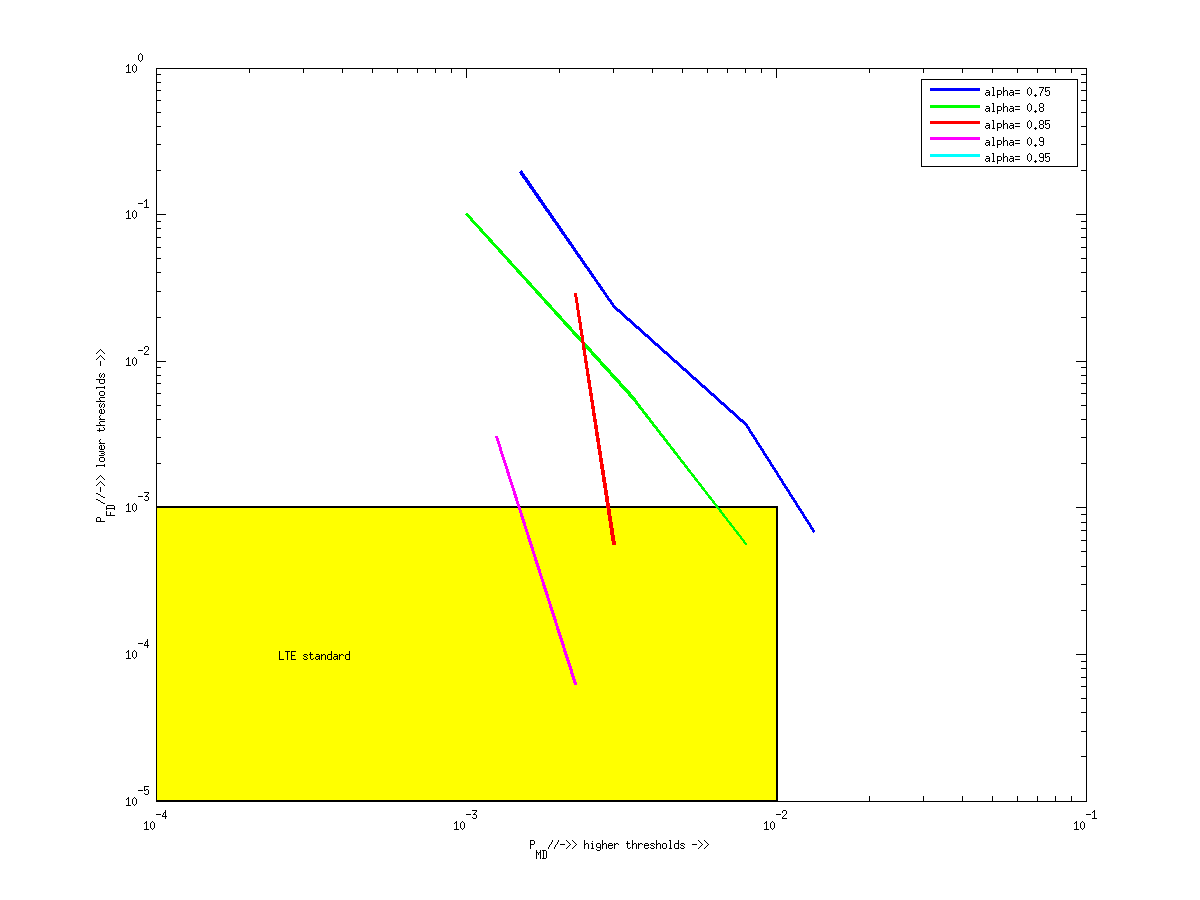}\caption{$P_{FD}%
=10^{-2}$ over $P_{MD}=10^{-2}$ for the 5G \textquotedblright
one-shot\textquotedblright\ random access with same parameters for the first
setting in Fig.\ref{fig:crach:ser}. It is observed that the LTE-A detection
(in yellow) can be achieved for certain $\alpha$. For the second setting in
Fig.\ref{fig:crach:ser} almost no detection errors are observed at overall
SNR=20dB}%
\label{fig:crach:pfd}%
\end{figure}

In our setting, a limited number of users is detected out of a maximum set
(here 10 out of 50). We assume that the delay spread is below 300 dimensions
of which only a set of 6 pathes are actually relevant. Each active user sends
1000 bits in some predefined frequency slot. This is uniquely achieved by
mapping the sequences to a slot. Hence, in the classical Shannon setting 50
users x (300 pathes + 1000 bits) = 65000 dimensions are needed while there are
only 24k available! The performance results are depicted in
Fig.\ref{fig:crach:ser} where we show show symbol error rates (SER) over the
pilot-to-data power ratio $\alpha$. Moreover, in Fig.\ref{fig:crach:pfd} we
depict false detection probability $P_{FD}$ (some user is detected while not
active) over missed detection probability $P_{MD}$ (user is active while not
detected). We observe that, although the algorithms might not yet capture the
full potential of this idea, reasonable detection performance can be achieved
by varying $\alpha$. In the 4G LTE-A standard a minimum $P_{FD}=10^{-3}$ is
required for any number of receive antennas, for all frame structures and for
any channel bandwidth. For certain SNRs a minimum $P_{MD}=10^{-2}$ is
required. It can be observed from the simulations that the requirements can be
achieved. Actually, compared to 4G LTE-A where the control signalling can be
up to 2000\% \cite{Wunder2014_COMMAG} of a single resource element the control
overhead is in the CS setting down to to 5\% (let alone the huge increase in
latency) in the best case!

A great challenge is the link between CS estimation and information-theoretic
rates. In \cite{WunderICC_2014} we have recently shown that the rate error per
subcarrier $i$ is lower bounded by%
\begin{equation}
\Delta r_{i}\left(  \alpha\right)  \leq\log\left(  1+\frac{m\cdot c_{2}%
(\delta_{2k_{1}})^{2}}{n}\left(  \text{SNR}\cdot\frac{\left(  1-\alpha\right)
\beta}{\alpha}+\frac{1}{\alpha}\right)  \right)  \label{eq:rate_error}%
\end{equation}
mimicking the curves' shape in the simulations. Here, the system parameter
$c_{2}(\delta_{2k_{1}})=4\sqrt{1+\delta_{2k_{1}}}/(1-(1+\sqrt{2}%
)\delta_{2k_{1}})$ depends only on the \emph{restricted isometry property}
$\delta_{2k_{1}}$ of the measurement matrix for $k_{1}$ sparse channels which
is "small" with high probability provided that $m\geq\mathcal{O}\left(
k_{1}\log^{5}(n)\right)  $ and $\beta>0$ is a constant not depending on $m,n$.

Finally, we mention that we have used a simple FDMA random access strategy for
the data which was recently discussed in \cite{Huang2013} along with another
more sophisticated slotted ALOHA access scheme employing a multiuser detector.
Yet, no sparsity is involved. Similarly in \cite{Popovski14} a random access
scheme called \emph{coded slotted ALOHA} is proposed investigating the
interaction of advanced multiuser detection based on CS and successive
interference cancellation. In both works the effect of channel/data estimation
errors and error propagation in the interference cancellation scheme is
crucial \cite{Popovski14} and must be carefully considered as it is done in
eq. (\ref{eq:rate_error}).

\subsection{Incoherent multiantenna receiver concepts}

The situation is even more intricate if both the sequence set and the channel
are unknown during sampling. For the purpose of exposition we consider the
single user/single antenna case and use the short--hand notation $\ast$ for
circular convolution. In such a blind reference model the signals to be
sampled are given, for example, as $h\ast x$ of two unknown but sparse vectors
$x$ and $h$ with $\lVert h\rVert_{0}\leq k_{1}$ and $\lVert x\rVert_{0}\leq
k_{2}$. The goals are (i) sample $h\ast x$ with minimal number of measurements
and (ii) determine from $h\ast x$ the pair $(x,h)$ up to indissoluble
ambiguities. Compressive sampling via $\Phi\in\mathbb{C}^{m\times n}$ of such
type of bilinear combinations in the strongly undersampled regime (the number
of measurements $m$ scales with $k_{1}+k_{2}$ although $\lVert h\ast
x\rVert_{0}\leq k_{1}k_{2}$) can be achieved under additional stability
assumptions, see \cite{walk:stability2013} and references therein. This
bilinear recovery problem can be lifted again to an LMR problem, since
circular convolution $h\ast x=B\left(  \text{vec}(hx^{T})\right)  $ can be
expressed as linear mapping $B$ applied on the (vectorized) rank--one matrix
$X=hx^{T}$. More precisely, $B$ is a suitable matrix with elements
$(B)_{i,(jk)}=\delta_{i,j\oplus k}$. A convex recovery algorithm not taking
sparsity into account follows then from \eqref{eq:lmr:l1min}. Such a nuclear
norm minimization has been considered by some authors for linearly random
encoded data $x$, see for example \cite{Ahmed2012_ARXIV} and further reference
therein (also related independent component analysis). Indeed, there it has
been shown that this approach is successful with overwhelming probability when
certain oversampling is used. It has to be expected that this situation
improves if sparsity will be taken into account and even undersampling can
then be used. For example, nuclear norm minimization could be $\ell_{1}%
$--penalized:
\begin{equation}
\min\,\lVert X\rVert_{\ast}+\lambda\left\Vert \text{vec}(X)\right\Vert
_{1}\quad\text{s.t.\ }\lVert y-\Phi B(X)\rVert_{2}\leq\epsilon
\label{eq:crach:blind}%
\end{equation}
for some regularization parameter $\lambda>0$. Yet, such multi-objective
convex methods will not scale better then the best separate optimization
\cite{Oymak2012}. Thus, it is of fundamental importance to find recovery
algorithms which (i) can operate at the optimal, additive ($k_{1}+k_{2}$)
scaling of measurements and (ii) can be easily extended to the case with
multiple interferers.

\subsection{Massive antenna regime}

The system in eq. (\ref{eq:crach:coherent}) in the non-coherent setting
culminates in a many to massive antenna design when the number of receive
antennas is scaled up. The standard setting for the massive MIMO case is the
flat fading case.

The standard setting for the massive MIMO case is the flat-fading case where
all $h_{q}^{p}$ contain a non-zero element at their first position only so
that we have $H_{p}=(h_{p}^{1},...,h_{p}^{q})=h_{p}^{T}$. The model can be
written as%
\begin{equation}
Y=\sum_{p=1}^{N_{t}}s_{p}h_{p}^{T}+Z
\end{equation}
It can then be shown that an SNR optimal detector is given by a \emph{singular
value decomposition} of $Y$ when sparsity is not involved. Very recently,
\cite{Cottatellucci2013_VTC} has linked this to the so-called \emph{pilot
contamination problem} in a multi-cell scenario using random matrix theory
showing that eigenvalue sets related to pilot signals in different cells
appear in fact in disjoint intervals so that they can be separated. Hence, by
this nonlinear detector the pilot contamination problem disappears, however,
at the cost of high receiver complexity involving "big" matrix decomposition,
of course. Notably, the general case involving sparsity and multipath fading
is an open problem and an important part of the future research agenda.

\section{Cloud Radio Access Networks}

\label{s:cloud}

In our target architecture, \emph{virtual base stations} located in the data
centers control a scalable number of nodes (and terminals) for which
transmission/reception is coordinated by sharing control information and/or
even messages (so-called \emph{cooperative designs} or \emph{coordinated
multipoint (CoMP)}). While it is appealing to exploit existing designs we
describe the short-comings of such approach in this section, and how sparse
signal processing concepts will enable such designs.

\subsection{State-of-the-art cooperative designs: A critical view}

A summary of state-of-the-art cooperative designs in multi-cell networks can
be found in \cite{Marsch2011}. However, existing cooperative designs, as they
are available at present, are not scalable, i.e., operational regions and
switching points depending on the available channel state information (CSI) as
well as suitable transitions between different technologies are not clearly
defined yet. A prominent coordinated beamforming technique is e.g.
interference alignment (IA) \cite{Jafar2011} which essentially aligns the
signal space so that multiple interferer appear in the same subspace. In some
scenarios, closed-form solutions are available while in others only iterative
solutions exists; rigorous convergence analysis or relevant stopping criteria
are still missing. In this context, a scalable control signaling architecture
in uplink (feedback) and downlink (feedforward) is a major requirement. In the
existing designs, each point-to-point link is treated separately using
orthogonal resources. Consequently, with densification of cells it becomes
virtually impossible to provide CSI to all coordinated nodes (or antennas).
For this small cell scenario, it is an intriguing idea to superpose and
compress the control signals suitably, and let the (virtual) base stations
recover their own channels by exploiting sparsity of channel profiles and
effectively coordinated nodes.

There is another major point which, we think, basically stems from the lack of
robustness in existing designs: Industrial field trials show rather
disappointing throughput gains of CoMP algorithms \cite{Irmer2011} far away
from the beforehand highlighted information-theoretic limits \cite{Marsch2011}%
, whereby the major limiting factor is again properly sharing CSI and other
overhead among cells. This so-called \emph{limited feedback problem} has been
greatly analyzed in \cite{Caire2010} (for multiuser MIMO) and recently in
\cite{Kerret2011} (for joint transmission) and \cite{Ayach2012} (for IA) in
terms of the rate distance $\Delta r_{i}$ of node $i$ to capacity subject to
some offset independent of SNR. Hence, these results essentially provide a
systems' degrees-of-freedom analysis, i.e., assuming \emph{infinite SNR
regime}. To be specific, let $p$ and $b=b(p)$ denote the SNR and feedback
budget (in bits/channel use) as a function of SNR, respectively, then the
per-node capacity degradation for any scaling in $p$ is (using order notation
$\mathcal{O}(\cdot)$):%
\begin{equation}
\Delta r_{i}(p,b)=\log(1+p\cdot2^{-\frac{b}{n_{t}-1}})+\mathcal{O}(1)
\label{eq:cloud:dof}%
\end{equation}
Very recent studies \cite{Schreck13:TWC:robustIA} indicate that these results
are fragile and that, in fact, the tradeoffs actually behave very different in
more practical regimes. It is shown that for \emph{any finite SNR point} $p$
and for any scaling in $b$ the per-node capacity degradation is:
\begin{equation}
\Delta r_{i}(p,b)=\mathcal{O}\left(  \log(1+p\cdot2^{-\frac{b}{2(n_{t}-1)}%
})\right)
\end{equation}
which actually \emph{doubles the required number of bits} compared to
(\ref{eq:cloud:dof}). Classical analysis falls short due to several reasons:
i) It assumes infinite SNR regime where achieving DoF is optimal. In this
operational regime interference mitigation instead of signal enhancement is
the primary goal. ii) It asserts that the transmitter can optimally allocate
rates while, in practice, the transmitter allocates rates according to the
available CSI and corresponding scheduler decisions (real versus ideal link
adaption). iii) The optimal scheduling decision is known a priori which is
unrealistic since limited feedback not only affects the choice of spatial
precoding but also user selection and resource allocation.

Obviously, the problem even worsens if a frequency-selective channel is
considered and also persists with alternative time domain quantization.
Altogether, the classical analysis renders the performance estimation overly
optimistic and it is safe to say that the relevant tradeoffs in a dense C-RAN
architecture are not yet well understood. This calls again for a highly
efficient control signaling architecture which then considers robustness from
scratch as outlined next.

\subsection{Towards a sparse architecture: Analog relay and linearly compress}

Let us introduce a control signaling architecture where the feedback
generating terminals receive the pilot signals from several nodes at different
receive antennas at once and act as simple \emph{relay and linearly compress}
nodes. Linear compressing means that instead of the complete set of pilot
measurements across time and spatial domain only a linear combination of them
is (separately) fed back (possibly in analog form). All cooperating nodes
collect these compressed control signals from the terminals and makes suitable
estimates of the channels.

There are several key features of this scheme: The pilot patterns from the set
of cooperating nodes are superposed and can be separated by advanced
processing exploiting CIR compressibility. The overhead does not scale with
the number of nodes joining, so that every node can join the cooperating set
if it wishes. Each cooperating node receives the same signal reflecting the
C-RAN architecture where the baseband processing of different nodes is in the
same place. Inherently by the linear compressing, only the most relevant
subset of nodes are effectively coordinated. Since this step is independent of
the used pilot pattern the base station can change it without informing the
terminals (so that it can be even random). Altogether, the base station takes
over all the processing which is affordable in a centralized C-RAN
architecture. From the terminals' point of view, neither do they need to know
which nodes are cooperating nor do they need to quantize any information of
the channels. By contrast, in simple time domain quantization, it is still
assumed that the paths of each antenna of each node is treated the same way
which becomes quickly inefficient in a highly dense C-RAN architecture where
only the most significant paths across all base stations should be compressed.
Moreover, the required communication resources should scale along this number
rather than all antennas and nodes. Finally, the terminals could even make use
of the same processing principles (only few might be active). Note that our
design is efficient and scalable such that it meets the number of essentially
unknown parameters (but \emph{not} the sampling theorem) which is typically
much smaller.

Mathematically, the scheme is expressed similar to the multiple access scheme
in eq. (\ref{eq:crach:coherent}). Denoting the number of nodes again by
$n_{t}$, the received signal at each of the cooperating nodes $y$ is:%
\begin{equation}
y=\Phi\left(  \sum_{p=1}^{n_{t}}S_{p}H_{p}+Z_{1}\right)  +z_{2}%
\end{equation}
Here, the inner bracket is what the terminal receives on multiple antennas
similar to eq. (\ref{eq:crach:coherent}). The main difference is the
additional relay component resulting in two different noise sources
$Z_{1},z_{2}$ which complicates the analysis and performance limits are not
known. Another significant challenge is to make such sparse designs robust
within the C-RAN. In \cite{Wunder2012_TWC,Schreck13:TWC:robustIA} we sketch a
robust design at finite SNR guided by the intuition to improve the overall
performance by suitable metrics (which capture the effects of e.g. scheduling)
and transmit/receive strategies (guaranteeing worst-case performance no matter
what the scheduling decision is) rather than simply approximating the channel.
Combining such robust design with the sparse signal processing paradigm
essentially requires to fully understand the tradeoffs between number of
measurements and the actual regarded (new) performance metrics. This is
actually far away from current achievements in the respective literature. A
good starting point is the rate expression in eq. (\ref{eq:rate_error}) though
with suitable estimates for the RIP parameters.

\begin{figure}[ptbh]
\centering\includegraphics[width=\linewidth]{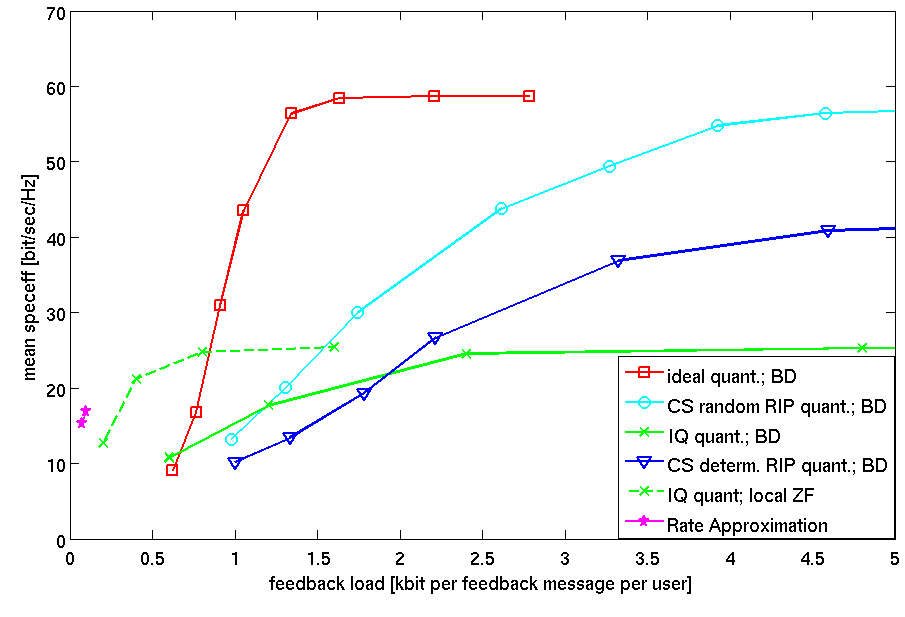}\caption{Mean
spectral efficiency of CoMP schemes over feedback load using IQ quantization
or CS based techniques: BD is block diagonalization, ZF is zero forcing and
Rate Approximation is the scheme from \cite{Wunder2012_TWC}}%
\label{fig:cran}%
\end{figure}

Let us mimic a 5G-like simulation scenario with LTE-A 4G parameters in Fig.
\ref{fig:cran}. Here, we compare the analog relay and linearly compress scheme
with standard IQ quantization, a genie scheme and the robust schemes described
in \cite{Wunder2012_TWC} in a LTE-A 4G setting. We consider 3 base stations
with $n_{t}=4$ transmit antennas located in 3 adjacent cells with 10 users
($n_{r}=1$) uniformly distributed over the network area (radius of 250 meter
around the center of the base stations). The physical layer is configured
according to 4G. The channels are modeled by the spatial channel model
extended (SCME) [15] using the urban macro scenario. It can be clearly
observed that the CS provides a much better scaling compared to simple IQ
quantization. Note, that we have not incorporated any sparse topology yet.

\subsection{Sparse prediction: Unexplored ground}

CSI aging and corresponding CSI mismatch due to asynchronous signaling is
another serious problem in a CRAN. In \cite{Caire2010} it was shown that
classical prediction is beneficial for zeroforcing beamforming (ZF) in
cooperative designs such that the systems are actually not fundamentally
interference limited. In case of just stale CSI \emph{retrospective
interference alignment} \cite{Mad2012} is another technique to be explored
within the developed sparse and robust context. Typically it is argued that
retrospective IAinterference alignment is actually outperformed by ZF over a
wide range of velocities \cite{Kobayashi2012a}. However, under the robustness
paradigm this might not be necessarily true as we proved that the performance
estimation for ZF is overly optimistic!

So far, the inherent sparsity is not considered in the prediction literature.
Clearly, sparsity provides additional structural information which can be
explored. The problems are intricate: the system (\ref{eq:crach:coherent})
might evolve on a complicated manifold not in a vector space. The simplest
approach is then to assume for the time evolution the same support of the path
coefficients for all times, which is reminiscent of block- or structured CS,
and to iteratively estimate the correct subspace and make a optimal prediction
on the estimated support next. Another approach is to let the process truly
vary on a sparse manifold where the simplest one is the union of all
$k$-sparse canonical subspaces in $\mathbb{C}^{n}$. However, so far, neither
specific models to enable e.g. advanced Kalman prediction nor any respective
algorithms have been developed.

The performance of classical short-time prediction algorithms suffer from the
poor granularity in the frequency domain and coarsely quantized complex CSI
due to limited feedback problem as well. Another key idea is, hence, to
develop a new framework based on sparsity where the sparse CIR is calculated
from simple real-valued measurements, cf. Sec.\ref{s:coding} for a strong
motivation of this assumption. Even though such an approach may not be
supported by a physical meaning in all cases, the new process is in any case
"slower" and therefore intuitively better predictable. Phase retrieval
algorithms can be used for such problems which we outlined in
Sec.~\ref{s:sparsity}. A special case are the Fourier measurements where the
problem is related to factorization theory. Alternative methods using
so-called \emph{symmetrized Fourier measurements} have been recently proposed
in \cite{walk:stability2013} which, subject to the sign, theoretically allow
stable reconstruction of the complex-valued CIR. Yet again, suitable
algorithms incorporating multiple antennas and nodes exploring the sparsity of
the CIRs are not known.

\section{Compressive Channel-Source-Network Coding}

\label{s:coding}

In this section we carry over the concepts to the terminal side following the
idea that sparsity may be already beneficial in the terminals' encoding process.

\subsection{Beyond the separation theorem}

5G networks, in particular the IoT component, will contain a large number of
relatively simple devices that need to operate at a very energy-efficient
level and with fairly limited memory. Hence these device can neither afford to
employ energy-hungry encoding or compression algorithms, nor can they support
high data rates. Thus power budget and communication rates become two main
design constraints for 5G networks. These constraints naturally suggest the
use of sensors that are built based on the compressive sensing paradigm. Using
compressive sensing devices (CSDs), one can avoid the need for power-consuming
compression algorithms, while still being able to transmit images of
sufficient quality without having to resort to transmitting images pixel by
pixel. Instead of recording data at a high resolution (which requires memory
and energy) and then throwing away most of the data during the compression
step (compression again requires energy), such CSDs collect data directly in
the \textquotedblleft compressive domain\textquotedblright. The computational
burden is shifted from the transmitter to the receiver. Yet, deploying CSDs in
a wireless network will impact the entire communication system design.
Understanding this impact and taking full advantage of it will be an important
aspect for efficient 5G network modeling.

A variety of interesting challenges arise in such \textquotedblleft
compressive sensor\textquotedblright\ situations. Unlike in the
\textquotedblleft classical\textquotedblright\ communication scenario, where
Shannon's separation theorem is one of the guiding principles for designing
communication systems (even though the idealized conditions on which the
separation theorem is based are hardly met in practice), we are now faced with
a different setup. The situation depends strongly on which constraints we put
on the complexity of the encoding step. For example, if we equip our simple
CSD with a fairly complex encoder, then we find ourselves back in a classical
communication scenario. This is due to the fact that in this case we could
incorporate both the reconstruction of the original signal from its
compressive samples and a standard source coding step into the encoding
procedure. But it is clear that such a complicated encoder would be
detrimental to our goal of having simple, power-efficient sensing devices.
Instead let us look at the other extreme, where we deal with a sensor that
does not permit any encoding at all. The only way to add redundancy (and thus
to add error protection) in that case is to increase the number of compressive measurements.

Following the CS paradigm such compressive measurements have to be slightly
redundant to allow for numerically efficient signal recovery (compared to the
number of measurements for an NP-hard reconstruction algorithm). For ideal CS
matrices the number of measurements $m$ for a $k$-sparse signal $x\in
\mathbb{C}^{n}$ has to be at least $\mathcal{O}(k \log(n/k))$. Hence, to
increase robustness, we may take $m > \mathcal{O}(k \log(n/k))$. Clearly, $m$
should depend on the SNR. A simple CS-based source coding scheme might look as
follows: An instructive numerical simulation illustrates the efficacy of this
simple scheme. Similar to~\cite{FM11} we consider the following setup:

\begin{itemize}
\item[(i)] \emph{Encoding:} The encoding step consists of taking $m \le n$
measurements of a signal $x\in C^{n}$ via $w=\Phi x$, where we assume that $x$
is $k$-sparse in a known basis $\Psi$.

\item[(ii)] \emph{Transmission:} The measured signal $w$ is transmitted
through a noisy channel. The received signal is $y=w+z = \Phi x+z$, where $z$
is AWGN.

\item[(iii)] \emph{Decoding:} At the receiver we attempt to recover $x$ via
the LASSO, i.e., by solving $\min\,\, \frac{1}{2} \|\Psi x\|_{1} +
\lambda\|\Phi x - y\|_{2}$.
\end{itemize}

Simulation examples can be found in~\cite{FM11}.


\subsection{Spatially correlated networks}

Let us now consider a sensor network where the source signals are sparse with
respect to some known basis and in addition are \emph{spatially correlated}.
Instead of decoding each signal individually, it is natural to try to exploit
the spatial correlation between the sensors. There are various ways to
accomplish this. In~\cite{FM11} an elementary sequential reconstruction
algorithm is proposed to exploit the spatial correlation between signals,
based on the fact that the difference between two sparse signals is again
sparse. A different, more flexible and robust approach consists of setting up
a low-rank optimization problem. By interpreting the signals as columns of a
data matrix $X$, the spatial correlation suggests that one may attempt to
find, among all the matrices consistent with the observed data, that with
minimal rank. We can try to improve upon this decoding procedure even further
by attempting to combine the powers of compressive sensing with those of
matrix completion. Hence we may consider an optimization problem that combines
the sparsity and the low-rank structures, as e.g.\ in \eqref{eq:crach:blind}.
While~\cite{Oymak2012} shows (in the noiseless case) that such mixed-convex
optimization approaches cannot lead to significant improvements over an
approach that optimally exploits one of the individual sparse properties,
numerical simulations nevertheless indicate that the improvements can still be
worthwhile from a practical viewpoint - in particular in the noisy case.
Recovery problems that combine sparsity and low rank as penalty function have
been proposed recently in the literature, but their careful use and
investigation in wireless communications is unchartered territory.

So far we have not made use of the fact that, unlike in the usual compressive
sensing scenarios, the coefficients of the signals we try to reconstruct
belong to some finite alphabet, such as e.g.\ QPSK. Hence, an important open
problem is to extend standard decision feedback equalization schemes to the
situations described in this section. How can we optimally exploit the
sparsity of the signals and the spatial correlation across sensors in a
decision feedback scheme? A possible approach might be to try to combine
recent advances in mixed integer optimization~\cite{LHK2012} with sparse
recovery methods.

Many variations of the theme are possible. We have ignored the role and effect
of quantization so far. Moreover, the sensors may allow for \emph{very simple}
encoding in addition to, or instead of, increasing the number of compressive
measurements. How does an analog of Shannon's separation theorem look like for
such a scenario?

\section{Embedded Security}

\label{s:security} Embedded security together with sparse signal processing
promoting simultaneously secrecy and reliability can fundamentally change the
way we approach the design of security, authentication and integrity
mechanisms, specifically in the IoT and the TI. In this section, we show that
sparse signal processing can be naturally incorporated within the concept of
embedded security and exhibits indeed a new degree of freedom in the design of
algorithms, naturally entailing new interesting tradeoffs considering
compressibility and secrecy.

\subsection{5G security challenges}

One example where current security solutions fall far short is the IoT due to
the \emph{scalability problem}: Nodes must be flexibly managed and distributed
in the network and asymmetric schemes used at the application layer are too
complex as well as too computational- and energy intensive for the typical
battery driven low-complex wireless transceivers of IoT devices. Symmetric key
schemes are 100 to 1000 times less complex ("lightweight security") but assume
a common secret key for the nodes so that that there is a \emph{key
distribution problem} instead. A clever detour of this problem is to
distribute unique keys already in the chip sets' manufacturing process (see
www.weightless.org). However, the overall security architecture still requires
Internet server access and lacks some flexibility.

An alternative concept is embedded security aiming at physical layer
integration of confidential services in wireless networks
\cite{SchaeferBoche13SPM}. Here, \emph{classical wiretap coding} can achieve
provably non-zero secrecy rates within the imposed (often idealistic) channel
model, where it is then impossible for an eavesdropper to extract any
information about the sent message from the overheard signal. The concept of
wiretap coding is somewhat detrimental to the scalability and low complexity
requirements in the IoT similar to the argument in Sec. \ref{s:coding}. A much
simpler method is to exploit the reciprocity and fading nature of CSI and to
establish a common secret between sender and transmitter from the CSI
measurements \cite{Gollakota2013}. Since keys are then automatically
installed, key distribution in these systems is easy to manage and requires no
user interaction; moreover beyond the spatial decorrelation length of antenna
elements such keys are virtually impossible to recover. Two major practical
problems occur though: i) imperfect reciprocity of CSI ii) insufficient
entropy of the generated keys due to static channels. In order to handle the
imperfect reciprocity, typically a so-called information reconciliation
procedure over a public channel is run, careful not to unveil any information
about the secret key bits. Imperfect reciprocity and insufficient entropy
depend on each other because "coarse" quantization and stronger codes improve
on key agreement rate but reduce the keys' entropy. Both aspect affords
additional control signaling, yet again bringing up the issue of reliability,
complexity, nodes' lifetime, and new security threats in the IoT.

Another example is the ultra-low latency requirement in the TI. Each and every
element of the communication and control chain must be optimized and,
obviously, fast authentication and secure communication is a "must" for the TI
then. We also emphasize the role of data integrity due to the high reliability
constraints, e.g. in the context of Industrial Wireless. Applying standard
security mechanisms on the application layer is not feasible. Moreover,
wireless channel secret key generation is highly limited in the rate of the
generated key bits, which is at most 44bit/s by today so that at least roughly
3s are required to generate e.g. a secure 128bit key \cite{Gollakota2013}. Too
slow for the TI!

To overcome such limitations, security shall be built in the compressive CSI
control signaling architectures developed in Sec.~\ref{s:randomaccess} and
Sec.~\ref{s:cloud} from scratch. Here, in contrast to the robust design by
taking sufficently many compressive measurements, the opposed direction is
taken to disguise the CSI, which shall be discussed next.

\subsection{Making security fast and scalable}

In our concept, a secret key is periodically generated from CSI and
acknowledged between transmitter and legitimate receiver. Let us assume that
information reconciliation is part of the "relay and compress" control
signaling architecture in Sec. \ref{s:cloud}. Then, since the legitimate
receiver can reliably recover the channel, reciprocity and key entropy is
preserved which advantageous in case of static channels. In addition, similar
to the discussion in Sec.~\ref{s:cloud} the transmitter can change the control
pattern without informing the receiver. Furthermore, since this is a full
duplex scheme the scheme is faster making it a candidate for the TI (standard
wireless channel secret key generation schemes typically run in half duplex
time division scheme). The catch is though that potentially the eavesdropper
can recover the key itself when he/she is able to collect all the control information.

In order to make such schemes workable in practice, additional measures should
be taken to ensure that an eavesdropper cannot reveal the message. Actually,
this sets a limit to the number of measurements publicly discussed such that
slightly erroneous or incomplete information about the control signaling
patterns as well as about the compressive measurements make it impossible to
extract the original messages. Clearly, one can think of an ocean of possible
communication protocols to improve on this line of thinking. Interestingly, by
using the wireless channel as the secret key source we have a new interesting
trade off between compressibility and secrecy: good compressibility means
small entropy in the key, hence longer observation times and vice versa.

To illustrate an example of such \textquotedblleft built-in\textquotedblright%
\ security we consider the following scenario. Let us assume that the relay
and compress scheme from Sec. \ref{s:cloud} is used in a point-to-point link
to inform the transmitter (Alice) about the channel of its legitimate receiver
(Bob). The purpose is to extract a common key from it in order to conceal the
information from an eavesdropper (Eve). Clearly, if Eve knows all the control
signalling it can recover the information from the measurements. In practice,
by the physical nature of wireless transmission, these measurements are only
available subject to some unknown phase shift for each measurement. The
simulation in Fig. \ref{s:cloud} then clearly indicates that Eve is not able
to recover the information while Bob can still get some reasonable
performance. The MSE performance is even worse for rank-one distortion. We
would like to emphasize that any analytical approach for the such
"perturbations" is not known to the best of our knowledge.

Allowing fast, efficient, and flexible key distribution are desirable
principles in the TI as well. However, to achieve fast authentication even
more advanced methods must be used ranging from wireless fingerprinting
\cite{Lai2011} which can include e.g. the individual sparsity patterns as well
as cooperative jamming approaches \cite{Gollakota2013}.

\begin{figure}[ptbh]
\centering\includegraphics[width=\linewidth]{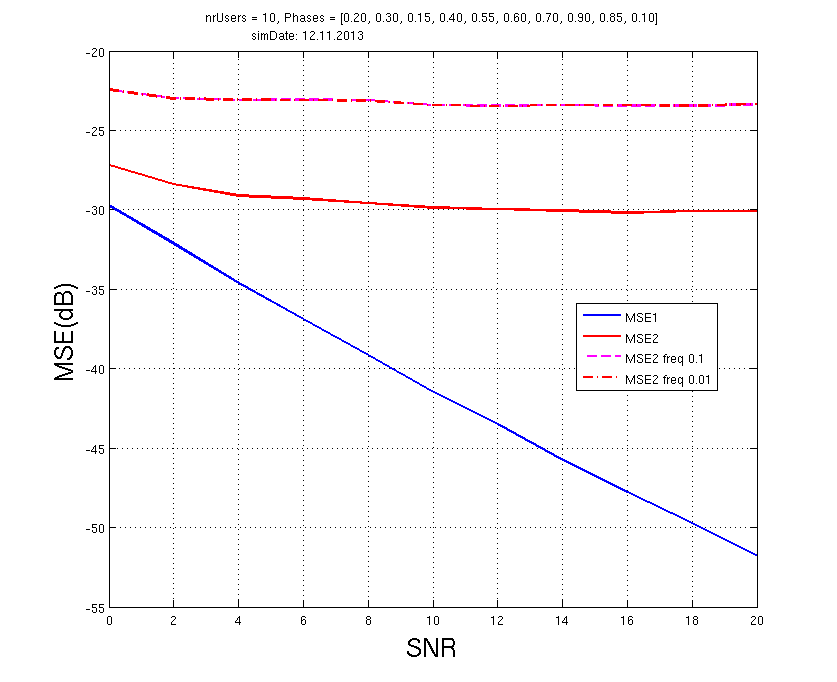}\caption{MSE
performance of the relay and compress scheme under "perturbations" for 1) Bob
having ideal measurements and 2) Eve having perturbed measurements with a)
phase errors and, worse, b) rank-one distortion}%
\label{fig:sec}%
\end{figure}

\section{Conclusion}

We have shown that sparse signal processing is indeed a viable source for an
innovative 5G system. To exploit the benefits fundamental research is required
addressing the many open questions regarding tradeoffs, performance limits,
algorithmic framework etc. We have only touched the surface of this research
agenda, where one of the many further fields of exploration is to include new
waveforms at the physical layer. We also emphasize that sparsity appears not
only as a physical reality, e.g. in the wireless channels, but also by design
of the network topology, traffic conditions etc. It is therefore an important
future task to measure the degree of sparsity in a system and adapt the
signaling architecture acccordingly.

\section{Acknowledgements}

The work of G. Wunder was supported by the 5GNOW project supported by the
European Commission under grant 318555 (FP7 Call 8). T. Strohmer acknowledges
partial support from the NSF via grant DTRA-DMS 1042939, and from DARPA via
grant N66001-11-1-4090. P. Jung was supported by DFG--grant JU 2795/2-1.

\noindent
\bibliographystyle{IEEEbib}
\bibliography{spm5g}

\end{document}